\documentstyle[aps,prl,tighten,floats,epsfig]{revtex}
\begin{document}
\draft
\widetext

\title{Relativistic calculation of the triton binding energy and its
implications}
\author{
Alfred Stadler$^{1,2*}$ and Franz Gross$^{1,3}$}
\address{
$^1$College of William and Mary, Williamsburg, VA 23185}
\address{
$^2$Centro de F\'\i sica Nuclear da Universidade de Lisboa, 1699 Lisboa Codex,
Portugal}
\address{
$^3$Thomas Jefferson National Accelerator Facility, Newport News, VA 23606}
\date{\today}
\maketitle
\begin{abstract}
First results for the triton binding energy
obtained from the relativistic spectator or Gross equation are reported. The
Dirac structure of the nucleons is taken into account. Numerical
results are presented for a family of realistic OBE models with off-shell
scalar
couplings.  It is shown that these off-shell couplings improve both the fits to
the two-body data and the predictions for the binding energy.
\end{abstract}
\vskip 1.5cm
\hskip 1.5cm
Preprint \#
WM--96--109,
TJNAF-TH-96-13,
CFNUL-96-5
%

\twocolumn


\narrowtext


The first realistic nonrelativistic calculations of the triton binding energy
were
completed in the 1970's\cite{T1}.  Later it was
shown that different methods gave the same results, and that the binding energy
could be calculated to an accuracy of a few keV by
considering all nucleon-nucleon ($NN$) partial waves up to $j=4$\cite{R1}.
Today, if three-body
forces (3BFs) are not considered, a small discrepancy of about 0.5-1.0 MeV
remains between the experimentally observed value of $-$8.48 MeV and values
obtained from realistic nonrelativistic $NN$ potentials.
State-of-the art calculations now include sophisticated 3BFs, and when their
strength is adjusted to give the
correct triton binding energy, an excellent value is also obtained for the
$^4$He binding energy (and to a lesser extent other light nuclei up to
$A\simeq7$)\cite{R2}.

However, relativistic effects should  make a contribution to the binding energy
at the level of several hundred keV.  Using a mean momentum of about 200 MeV
(consistent with nonrelativistic estimates) we expect to see corrections of the
order of $(v/c)^2\simeq (p/m)^2 \simeq 4\%$.  If this is 4\% of the binding
energy, then it amounts to about 300 keV.  However, if relativity has a greater
effect on the attractive $\sigma$ exchange
part of the force (as it does in nuclear matter calculations using the
Walecka model) then we might obtain an effect 10 times larger.

The importance of this problem has been recognized, and  relativistic effects
have been estimated using a separable kernel in the Bethe-Salpeter
equation\cite{RT}, assuming minimal relativity in the Blankenbecler-Sugar
equation\cite{MSS}, and by adding corrections of first order in $(v/c)^2$ to
the
Schr\"odinger equation\cite{Urbana}.  All of these calculations include some
contributions coming from relativistic kinematics, but none treats the full
Dirac structure of the nucleons, or investigates effects which might arise from
a realistic relativistic treatment of the $NN$ {\it dynamics\/}.

The purpose of this letter is to
present the first numerical
calculations of the triton binding energy obtained from the
manifestly covariant three-body spectator
(or
%
Gross) equations for three
identical spin 1/2 particles, and to discuss the implications of these
calculations.  Some preliminary results were reported in conference
proceedings\cite{conf}.


The three-body spectator equations were first introduced and applied to scalar
particles in 1982\cite{G1}, and then extended to the case of
three spin 1/2 particles in lectures given at the University of Hannover
soon afterward. Recently, a more tractable form for the equations has been
developed, and a full derivation of the equations will be published
elsewhere\cite{SGF}.  In this letter we describe only a few of their features
briefly.

In the absence of 3BFs the three-body scattering amplitude is obtained from a
sum
of all successive two-body scatterings.  Because the three particles are
identical,
each two-body scattering differs from the others only by a permutation, and
they
can therefore all be summed by one operator equation of the form
\begin{equation}
| \Gamma^1 \rangle  =2 M^1 G^1  {\cal P}_{12}  | \Gamma^1 \rangle\, ,
\label{Eq1}
\end{equation}
where $|\Gamma^1\rangle$ is a vertex function describing the contribution
to the bound state from all processes in which the 23 pair was the last to
interact (with particle 1 a spectator), the two-body amplitude $M^1$ describes
the scattering of the 23 pair,
$G^1$ is the propagator for the 23 pair, and $P_{12}$ is a permutation
operator interchanging particles 1 and 2.  (The factor of 2 comes from the
contribution of $P_{13}$ which equals the one of $P_{12}$.)

The three-body spectator equations have the same structure as
(\ref{Eq1}), but incorporate the additional feature that the spectator
is restricted to its positive energy mass-shell in all intermediate
states.  With the conventions implied above,
consistency also requires that particle 2 be on-shell, so that two
particles are always on-shell.  We think of
these constraints as a reorganization of Eq.~(\ref{Eq1}) which will, in some
cases, improve its convergence.
The constraints are manifestly covariant, and lead to the following equation
\begin{equation}
| \Gamma^1_2 \rangle  =  2 M^1_{22} G^1_2 {\cal P}_{12}
| \Gamma^1_2 \rangle  \, , \label{Eq2}
\end{equation}
where the lower index labels the second on-shell particle.  Hence
only particle 3, the (unique) particle which has just left one interaction and
is
about to enter another one, is off-shell in Eq.~(\ref{Eq2}).

To reduce Eq.~(\ref{Eq2}) to a practical form, we take matrix elements of
the operators using three-particle helicity states similar to those
defined by Wick\cite{Wick}.  Both $\rho$-spin states (where $\rho=+$ is the
$u$
spinor positive energy state and $\rho=-$ is the $v$ spinor negative
energy state) of the off-shell particle must be treated.  The three-body states
will be written in the abbreviated form
$| J 1 (23)\rho \rangle$, where $J$ is the total angular
momentum of the state, $\rho$ the $\rho$-spin of the off shell particle,
$1=\{q,\lambda_1\}$ (where $q$  and $\lambda_1$ are the magnitude of the
three momentum and the helicity of the spectator in the three-body c.m.), and
$(23)=\{\tilde{p},j,m_j,\lambda_2,\lambda_3\}$ (where $\tilde{p}$ is the
magnitude of the relative three momentum of the 23 system, $j$ and $m_j$ are
the
angular momentum of the pair and its projection in the direction of ${\bf
q}$, and $\lambda_2$ and $\lambda_3$ are the helicities  of particles 2
and 3, {\it all defined in the rest frame of the 23 pair}).  We will suppress
all
isospin indices.  Using this notation, the final form of the three-body
spectator
equation  for $\Gamma^1$ is

\begin{eqnarray}
&&\langle J 1 (23)\rho | \Gamma^1 \rangle =
\sum_{j'm'}\!
\sum_{{\lambda''_2 \lambda''_3\,\rho''}
\atop {\lambda'_1 \lambda'_2 \lambda'_3\,\rho'} }
\int_0^{q_{\hbox{{\tiny crit}}}} {q'^2 dq'}{m\over
E_{q'}}
\int_0^\pi d\chi \sin\chi \;
 \nonumber\\
&&\quad\times
\langle j (23)\rho | M^1| j (2''3'')\rho''\rangle
\,{m\over E_{\tilde{p}''}}\, g^{\rho''}(q,{\tilde p}'')\nonumber\\
&&\quad\times{\cal P}_{12}^{\rho''\rho'}[1(2''3''),1'(2'3')]
\,{m \over E_{\tilde{p}'}}\,
\langle J' 1' (2'3')\rho' |\Gamma^1 \rangle \, ,  \label{Eq6.1}
\end{eqnarray}
%
%
where ${\cal P}_{12}^{\rho''\rho'}[1(2''3''),1'(2'3')]$ is the matrix element
of the
permutation operator, given below, and $g^{\rho}(q,{\tilde p})$ the
propagator of the off-shell particle in different $\rho$-spin states
\begin{equation}
g^+(q,{\tilde p}) = {\displaystyle{1\over 2E_{{\tilde p}} -W_q}} \, ,
\qquad
g^-(q,{\tilde p}) = -{\displaystyle{1\over W_q }} \, .
\label{Eq6.2}
\end{equation}
Because four-momentum is conserved in the relativistic formalism,
the mass $W_q$ of the 23 pair depends on $q$,
\begin{equation}
W_{q}^2=M_t^2+m^2-2M_tE_q\, , \label{wq}
\end{equation}
with $E_q=\sqrt{m^2+{\bf q}^2}$.
Note that Eq.~(\ref{Eq6.1}) includes a sum
over intermediate helicities and angular momentum quantum numbers, and an
integration over the internal spectator momentum $q'$ and the angle
$\chi$ between the directions of ${\bf q}'$ and ${\bf q}$. The integration over
$q'$ has been
limited to the finite
interval $[0,q_{\hbox{{\scriptsize crit}}}]$, where
$q_{\hbox{{\scriptsize crit}}}$ is the root of the equation
$W_{q_{\hbox{{\scriptsize crit}}}}=0$.  At this critical spectator momentum
 (equal to $\simeq4m/3\simeq1200$ MeV), the two-body subsystem is
recoiling at the speed of light and the relativistic effects are enormous!
Contributions for $q'>q_{\hbox{{\scriptsize crit}}}$ are very small, and
come from two-body states with {\it spacelike\/}
four-momenta. It seems sensible to simply neglect the region $q'\ge
q_{\hbox{{\scriptsize crit}}}$ and set the three-body amplitudes to zero there.
As it turns out, the solutions
go smoothly to zero as $q\to q_{\hbox{{\scriptsize crit}}}$ anyway,
so we may impose the condition that they are zero beyond this point without
making the amplitudes discontinuous in $q$.

Exchanging particles 1 and 2 implies that particle 2 becomes the
spectator and now its momentum and helicity must be expressed in
the c.m.\ frame of the three-body system, while the variables of
particles 1 and 3 must be expressed in the rest frame of the 13 pair.
Boosting from one frame to another introduces Wigner rotations of both the
single particle and two-body helicities.  The final result for the
permutation operator is
%
\begin{eqnarray}
&&{\cal P}_{12}^{\rho''\rho'}[1(2''3''),1'(2'3')]=
(-1)^{m-\lambda_1+ \lambda'_3}
\sqrt{2j+1}\sqrt{2j'+1} \nonumber\\
&&\qquad \times d^{(J)}_{m-\lambda_1,m'-\lambda'_2}(\chi)
d^{(j)}_{m,\lambda''_2-\lambda''_3}(\tilde{\theta}'')
\,d^{(j')}_{m',\lambda'_1-\lambda'_3}(\tilde{\theta}')
\nonumber\\
&&\qquad\times
\,d^{(1/2)}_{\lambda_1\lambda'_1}(\beta_1)\,
d^{(1/2)}_{\lambda''_2\lambda'_2}(-\beta_2)\,
{\cal N}^{\rho''\rho'}_{\lambda''_3 \lambda'_3}
(q,q',\chi)\, ,
\end{eqnarray}
where the functions $d^{(1/2)}_{m_1,m_2}(\beta)$ are the Wigner rotation
matrices, and
${\cal N}^{\rho''\rho'}_{\lambda''_3\lambda'_3}(q,q',\chi)$ describes
{\em exactly}
the Wigner rotations of the off-shell particle 3, as well as the
nontrivial
matrix elements between the different $\rho$-spinors
$u$ and $v$ of particle 3 as they appear in the rest frames of the 23 pair and
the 13 pair.


We have solved Eq.~(\ref{Eq6.1}) numerically for a variety of realistic
$NN$ models.  The two-body amplitudes obtained for all of these models result
from an exact solution of the two-body spectator equation, as described in
Ref.~\cite{GVOH}, and are therefore fully consistent with the three-body
equations.

These models will be described in detail elsewhere.  Briefly, they
are all one-boson exchange (OBE) models with a kernel composed of the exchange
of 6 commonly used bosons: the $\pi, \eta, \sigma, \delta, \omega,$
and $\rho$.  The parameters of each model were determined by fitting to
the $NN$ phase shifts below 350 MeV and to deuteron properties.

In all cases the
following pion coupling was used:
\begin{eqnarray}
g_\pi\Lambda_\pi=&& g_\pi \left[ \gamma^5 -
{\displaystyle{ \nu_\pi \over 2m}}
\left[\left(m-\setminus\!\!\! p'\right)\,\gamma^5 +
\gamma^5\, \left(m-\setminus\!\!\! p \right)\right] \right]\nonumber\\
=&& g_\pi \left[
(1-\nu_\pi )\gamma^5 +
{\displaystyle{ \nu_\pi \over 2m}}
\gamma^5\,\setminus\!\!\!\! q  \right] \, ,
\end{eqnarray}
\noindent where $p$ and $p'$ are the four momenta of the incoming and
outgoing nucleons, and the couplings proportional to $\nu_\pi$ do not
contribute if the nucleons are on-shell.
In this
family, we fixed $g_\pi^2/4\pi=13.34$ and chose $\nu_\pi=1$, giving the
conventional pseudovector pion coupling with large off-shell effects.

A particular feature of these models, and a central point of this letter, is
that they
 {\it also\/} include phenomenological
scalar $\sigma$ (with $I=0$) and $\delta$ ($I=1$) exchanges with
off-shell scalar-nucleon-nucleon ($sNN$) couplings of the form
\begin{eqnarray}
g_s\Lambda_s(p',p)=&& g_s \left[ 1 -
{\displaystyle{ \nu_s \over 2m}}
 \left(m-\setminus\!\!\! p' +m
-\setminus\!\!\! p \right) \right. \nonumber\\
&&\qquad \left.  +{\displaystyle{ \kappa_s \over 4m^2}}
 \left(m-\setminus\!\!\! p'\right)\left(m
-\setminus\!\!\! p \right)\right] \, . \label{eq10}
\end{eqnarray}
The vertex $\Lambda_s(p',p)$ given in Eq.~(\ref{eq10}) is the {\it most general
form\/} the $sNN$ vertex can take, but as far as
we know the off-shell scalar couplings which depend on $\nu_s$ and $\kappa_s$
have never been studied previously.  The family of models discussed here has
$\kappa_s=0$ and values
of $\nu$ varying from $0\to2.6$, where
\begin{equation}
\nu_\sigma=-0.75\,\nu\qquad\quad
\nu_\delta=2.60\,\nu\, . \label{eq2}
\end{equation}
We will see that these couplings proportional to $\nu$ are extremely important.

\begin{figure}[t]
\vbox{
\centerline{\hbox{\epsfysize=3in \epsfbox{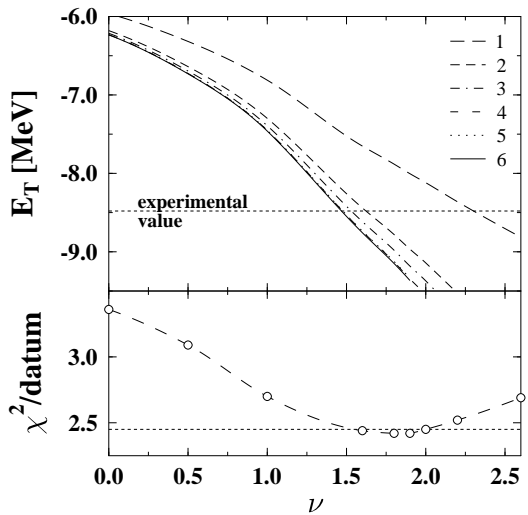}}}
\vspace*{-5mm}
\caption{Triton binding energy $E_t$ for specific values of $j_{max}$
(upper panel) and $\chi^2$ for the fits
to the two-body data (lower panel)
versus the scalar meson off-shell parameter $\nu$ defined in the text.  The
curves in both panels are smooth interpolations through the actual
calculations.
The lower panel also includes the line $\chi^2=2.45$ for reference.}
\label{fig1}
}
\end{figure}

The results of our calculations are summarized in Fig.~1. The lower panel of
Fig.~1 shows how $\chi^2$ for the
fits to the two-body data varies with $\nu$ for this family of OBE models.
Fits
were done for values of $\nu$ = 0, 0.5, 1.0, 1.6, 1.8, 1.9, 2.0, 2.2, and 2.6,
and the dashed curve smoothly interpolates these individual cases.
We emphasize that {\it each of these models with different values of $\nu$ are
realistic\/} in the sense that for each case OBE parameters (13 in all) were
adjusted to give the best possible fit to the $NN$ data below 350 MeV.
Although the 13
parameters differ only slightly from case to case,
the models are not quite equivalent. The
figure shows that there is a significant variation in the quality of the fit;
the best models lie in the region $1.5\le\nu\le2.0$ (all with $\chi^2\le 2.45$
as shown in the figure).  This probably rules out the model with $\nu=0$ (for
example).
We conclude that the introduction of these $\nu$-dependent couplings
significantly improves the fit to the two-body data and that the implicit
choice of $\nu=0$ made in previous work is not optimal.

The upper panel of Fig.~1 shows the variation of the three-body binding energy
with $\nu$.
The rapid dependence of the binding energy on $\nu$ is rather striking.  An
increase in $\nu$ from 0 to 1.6 changes our prediction from $-6.24$ to $-8.76$
MeV,
and a value in good agreement with experiment would be obtained for
$\nu\simeq1.5$, still in the range of $\nu$'s which give the best fit to the
two-body data.

The panel and Table 1 (for the cases with smaller $\nu$)
also show how the binding energy converges as the number
of three-body partial waves, characterized by the highest included pair angular
momentum $j_{max}$,
increases.  Because of the large increase in the predicted values as the number
of channels increases from 28 ($j_{max}=1$) to 52 ($j_{max}=2$), we were
concerned about the convergence of the three-body calculations and studied
it in detail.
We carried the
calculations all the way to $j_{max}=6$ with 148 channels. We find that the
individual contributions from channels with odd $j$ tend to cancel while those
from channels with even $j$ are all attractive. Thus, the steps from even to
odd
$j_{max}$ are small compared to those from odd to even $j_{max}$.  From a
detailed study of the individual contributions we estimate that the results are
converged to about 1 keV for $\nu=0$ and to
about 5 keV for $\nu =1.6$.

We conclude that the best of the two-body models examined so far yield a
three-body
binding energy from about $-$8.5 to $-$9.5 MeV.  In subsequent work we will
display the dependence of these results on the boosts, the negative $\rho$-spin
states, and other relativisitc effects, and we will study additional two-body
models.  Here we will discuss the origin and implications of the $\nu$
dependence which we have observed.

To understand why the binding energy is so sensitive to $\nu$,  we may look at
the half off-shell Born amplitude for scalar exchange (i.e. the amplitude with
{\it
one} nucleon off-shell).  For the positive $\rho$-spin sector, we have
\begin{eqnarray}
{\cal V}_s&&={g^2_s\left\{\overline{u}({\bf p}')
\left[ 1 - {\displaystyle{ \nu_s \over 2m}}
 \left(m-\setminus\!\!\! p \right) \right]
u({\bf p})\right\}\;
\left\{\overline{u}(-{\bf p}')
u(-{\bf p})\right\}\over m_s^2-(p'-p)^2}\nonumber\\
&&\simeq V_s\left[1- \nu_s{\displaystyle{ 2E_p-W \over 2m}} \right]=C_sV_s\, ,
\end{eqnarray}
where $V_s$ is the usual scalar potential obtained from such a reduction when
$\nu_s=0$, $p=(W-E_p, {\bf p})$ is the momentum of the
off-shell particle, $W$ is the energy of the two-body
system in its rest frame, and we have ignored the lower components of the
Dirac spinors in carrying out the reduction.  The effect of the $\nu_s$
dependence is to multiply the scalar potential by the factor
$C_s=[1-\nu_s(2E_p-W )/ 2m]$.
In applications to two-body scattering, the $\nu$-dependent term is a small
correction
with a sign depending on the energy, but in the three-body bound state it
is always positive. Assuming an average nucleon momentum of about
200 MeV gives roughly a 10\%
variation over the range of $\nu$ from 0 to 2.  The observed variation of
about 4 MeV
over this range is explained therefore if the average
strength of the $\sigma$-exchange potential is about 40 MeV,
which is the right order of magnitude.  This shows how the large variation in
binding energy which we observe can be explained by a ``small'' relativistic
effect.

\begin{table}
\caption{Absolute values of the triton binding energies in MeV. The first row
is the result when only $^1S_0$ and $^3S_1$-$^3D_1$ positive energy channels
are included.
The other rows show results obtained when all channels with two-body angular
momentum $j\le j_{max}$ are included.  The total number of three-body channels
in each case is N.}
\begin{tabular}{ l r  c c c c }
 $j_{max}$  & N & \multicolumn{4}{c}{coupling parameter $\nu$}\\
 &  & 0.0 & 0.5 & 1.0 & 1.6  \\
\tableline
 $1^+$ & 5 & 6.003 & 6.345 & 6.850 & 7.769  \\
\tableline
 1 & 28 & 5.963 & 6.318 & 6.812 & 7.652 \\
 2 & 52 & 6.180 & 6.639 & 7.299 & 8.441 \\
 3 & 76 & 6.214 & 6.695 & 7.393 & 8.615 \\
 4 & 100& 6.232 & 6.726 & 7.452 & 8.740 \\
 5 & 124& 6.233 & 6.726 & 7.452 & 8.736  \\
 6 & 148& 6.235 & 6.731 & 7.461 & 8.757

\end{tabular}
\end{table}

An OBE model with off-shell couplings has a
very rich structure.  For example, consider two successive interactions
of a scalar meson with a single nucleon.  The vertex function (\ref{eq10})
contains the operator
$m-\setminus\!\!\! p$ which  is just the inverse of the nucleon propagator,
so that it can remove the nucleon propagator and contract the two interaction
vertices to a single vertex describing the emission of two mesons from a single
point.
If the
two mesons emerging from this point are coupled to a second nucleon
they generate a triangle or a bubble diagram.  These diagrams are two-boson
exchange terms similar to those (involving pions) which would emerge from a
nonlinear sigma model.  Alternatively, if these two mesons couple to two {\it
different\/} nucleons, they generate diagrams usually associated with
three-body forces.  It is easy to generalize this result: {\it an OBE model
with
off-shell couplings is equivalent to another OBE model without these couplings,
but with an additional specific family of $N$-boson exchange diagrams and
$N$-body forces}.

We conclude with two observations.  The discovery that off-shell scalar
couplings play an important role in both improving the description of two-body
data and in predicting three-body binding energies would only have been
possible in the context of a relativistic formalism closely connected to an
effective field theory. The fact that these couplings are equivalent to a
strong energy dependence in the context of nonrelativistic theory is precisely
the reason they could not have been discovered there; nonrelativistic
potentials are supposed to be energy independent.  In the context of an
effective  field theory, however, these are a natural and legitimate extension
of the simplest assumption about the spin structure of the $sNN$ vertex.
The most general $sNN$ vertex was given in Eq.~(\ref{eq10}) above, and
can have only three different spin couplings.
Once the third term depending on $\kappa_s$ in (\ref{eq10}) is studied, all of
the possibilities will have been exhausted.  In this way an effective field
theory is tightly constrained, even if some of its interactions are strongly
energy dependent in a nonrelativistic context.

We believe that this way of looking at  dynamics may very well be the most
significant contribution to come from relativistic methods.  The traditional
arguments suggesting that relativistic effects are very small refer to
relativistic {\it kinematics\/} only.  As Eq.~(\ref{eq10}) illustrates,
relativistic {\it dynamics\/} provides a new way to study nuclei, even at low
energies.


One of us (FG) acknowledges the hospitality of Peter Sauer and the
University of Hannover, where the initial idea to persue this program began,
and the theory group at CFNUL where this letter was written.  Early
work was partially supported by a grant from NATO.
The major part of this work could not have been done without the support
of the Department of Energy
under grant \#DE-FG05-88ER40435 which is gratefully acknowledged. In
addition, one of us (AS) thanks JNICT for support under contract
\# BCC/4394/94.
Numerical calculations were performed at NERSC (Livermore),
TJNAF, and at CFNUL.

\end{document}